\documentclass[aps,prl,groupedaddress,preprint,showkeys]{revtex4}
\usepackage{graphicx}

\begin{document}

\title{Structural and dynamic features of liquid Si under high pressure
above the melting line minimum }

\author{T. Demchuk$^1$, T. Bryk$^{1,2}$, A.P. Seitsonen$^{3}$
}

\affiliation{
$^1$Institute for Condensed Matter Physics of the
National Academy of Sciences of Ukraine,
1 Svientsitskii Street, UA-79011 Lviv, Ukraine\\
$^2$ Institute of Applied Mathematics and
Fundamental Sciences, Lviv Polytechnic National University, UA-79013
Lviv, Ukraine\\
$^3$ D\'{e}partement de Chimie, \'{E}cole Normale Sup\'{e}rieure,
24 rue Lhomond, F-75005 Paris, France\\
}

\date{\today}

\begin{abstract}
We report an {\it ab initio} simulation study of changes in structural and dynamic properties 
of liquid Si at 7 pressures ranging from 10.2 GPa to 24.3 GPa along the isothermal line 1150~K,
which is above the minimum of the melting line. The increase of pressure from 10.2 GPa to 
16 GPa causes strong reduction in the tetrahedral ordering of the most close neighbors. 
The diffusion coefficient shows a linear decay vs drop in 
atomic volume, that agrees with theoretical prediction for simple liquid metals, thus not 
showing any feature at the pressures corresponding to the different crystal phase boundaries.
The Fourier-spectra of velocity autocorrelation function shows two-peak structure at 
pressures 20 GPa and higher. These characteristic frequencies correspond well to the 
peak frequencies of the transverse current spectral function in the second pseudo-Brillouin zone.
Two almost flat branches of short-wavelength transverse modes were observed for all the studied
pressures. We discuss the pressure evolution of characteristic frequencies in the longitudinal 
and transverse branches of collective modes.
\end{abstract}


\maketitle

\section{Introduction}
The behavior of condensed matter under high pressure is a fascinating field of studies which 
is important not only for fundamental knowledge of the material stability, but also for 
understanding the geophysical processes and for search for exotic structures and exotic properties,
which can be manifested only at extreme conditions. As an example, a new exotic mechanism of 
metal-nonmetal transition at high pressures via localization of
electrons in the interstitial areas, which was suggested in 2008 \cite{Rou08}, 
triggered a number of experimental observations of this phenomenon, in particular in Li \cite{Mat09}
 and Na \cite{Ma09}. The obtained results changed absolutely the existing knowledge about the 
dependence of the conduction bandwidth of metals on the interatomic separation. Moreover, {\it 
ab initio} atomistic simulations, performed for molten Li and Na, revealed an exotic tetrahedral 
ordering in liquid Lithium \cite{Tam08} and unusual electronic transitions in molten Na \cite{Rat07}.

Liquid Si is one of the most interesting systems with partial covalent bonding, which can reveal 
liquid-liquid transitions in the supercooled state \cite{Sas03,Jak07,Jak09,Gan09,Bey10}, which still 
is of great interest for simulation studies\cite{Zha17,Rem18,Bon18} for understanding different 
features of structural and dynamic properties as well as nucleation in liquid Si. Collective dynamics
in liquid Si was much less studied. In 2006 Delisle et al \cite{Del06} performed an 
orbital-free {\it ab initio} simulation study of liquid Si in the pressure range 4-23 GPa and 
for the temperatures 50~K above the melting line in order to compare their data to X-ray diffraction
experiments by Funamori and Tsuji\cite{Fun02}. In \cite{Del06} the authors reported dynamic structure 
of liquid Si, longitudinal and transverse currrent spectral functions as well as dispersion of 
collective excitations.

Collective dynamics of liquid metals especially on the spatial scales comparable to interatomic 
distances is not really clearly understood. In particular there were several reports on analysis of 
experimental dynamic structure factors obtained from inelastic X-ray scattering, in which the 
authors suggested nonzero contributions from transverse collective excitations to dynamic structure
factors\cite{Hos09,Gio10,Hos13}. Later on in {\it ab initio} molecular dynamics (AIMD) simulations of
liquid Li \cite{Bry15} revealed that at high pressures the transverse current spectral functions 
showed a two-peak structure, which became more pronounced with increasing pressure. Later on similar 
findings were observed even at ambient pressure in liquid Tl\cite{Bry18}. 

The main task of this study is to find similar features in single-particle and collective dynamics 
of liquids Si along an isothermal line above the minimum of melting line. The region around the 
minimum of melting line in liquid metals is of special interest because of possibility to observe 
there liquid-liquid transitions. The isothermal line on the phase diagram was chosen because of the 
need to have in simulations systems with the same mean thermal velocity of particles but with 
different pressures. We will calculate the static properties of liquid Si in the chosen region on
the phase diagram and will study the single-particle and collective dynamics. The next section 
reports the details of our AIMDs. The third section contains the results and discussion and the 
last section contains conclusion of this study.

\section{{\it Ab initio} simulations}

We simulated liquid Si by the VASP package using a system of 300 particles in
a cubic box subject to periodic boundary conditions. Seven pressures in the 
range 10.2 GPa-24.3 GPa at the temperature of 1150~K were studied in NVT 
ensemble. The isothermal line of 1150~K is above the minimum of the melting 
line in Si\cite{Bun64,Kub08} (see Fig.1) and above the region of several
crystal structures of Si. The initial system at the
lowest pressure was obtained from classical simulations of liquid Si, while 
compressed liquid Si was sequentially obtained from the last configuration of
lower density by rescaling the coordinates and the box size with subsequent 
equilibration over 4 ps. The time step was 2 fs, and the length of equilibrium
production runs were not less than 20 000 configurations for each pressure.

For electron-ion interaction we took the PAW potentials\cite{Blo94,Kre99}, which allow to recover
correct nodal structure of wave functions as well as correct
distribution of electron density in the core region. The exchange-correlation 
functional was taken in GGA-PBE\cite{Per96} form. Because of the rather large 
size of the simulated system we took for the sampling of the electron density 
only the Gamma point in the Brillouin zone.

The smallest wave number in our simulations was in the range from 
0.3815\AA$^{-1}$ for the lowest pressure and up to 0.3982\AA$^{-1}$ for the 
highest pressure. For calculations of the $k$-dependent quantities (static
correlators and time correlation functions) we sampled all possible $k$-vectors
with the same absolute value, and averaged the calculated $k$-dependent quantity
over all possible directions of these $k$-vectors.

\begin{figure}
\includegraphics[width=0.7\textwidth]{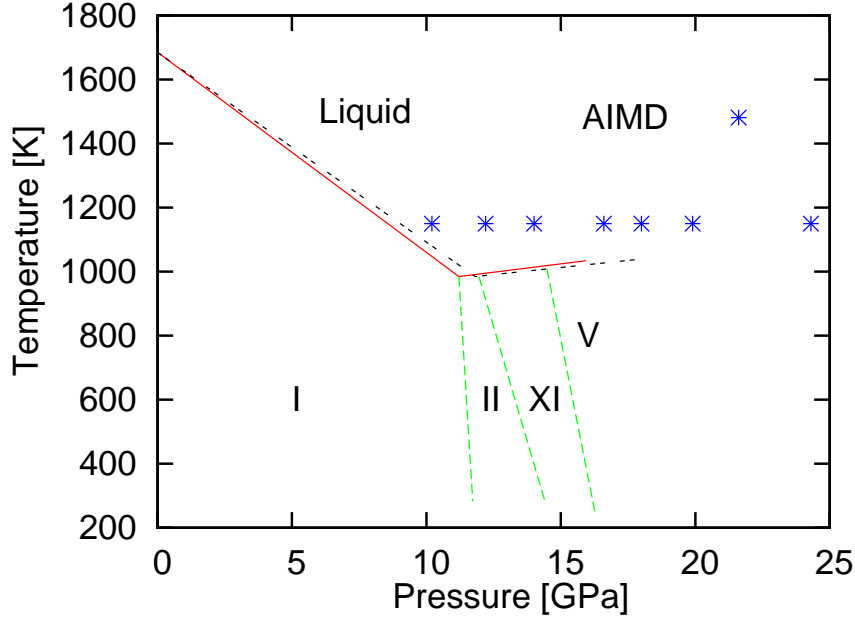}%
\caption{Phase diagram of silicon with the thermodynamic points reported 
in this study (asterisks). The melting lines for Si were taken from Refs.[21,22].
}
\label{ph_diagr}
\end{figure}
\section{Results and discussion}

The standard analysis of atomic structure of the disordered systems can be performed via pair
distribution functions (PDF), which are shown for the studied pressures along the isothermal line 
of 1150~K in Fig.2a. The main feature of the pressure dependence of PDFs in the range 10.2-24.3 GPa 
is the stable position of the first peak of PDF at 2.44\AA. Another interesting feature is almost
flat region between 3.1\AA~ and 4.2\AA~ of the PDF at the lowest studied pressure, which reduces 
with increasing pressure and transforming into well-defined minimum of PDF at 3.29\AA~ 
for the pressure 24.3 GPa.  The second maximum of PDF slightly changes its location from 
4.99\AA~ at 10.2 GPa to 4.78\AA~ at 24.3 GPa. We compared in Fig.2b the calculated PDF at the 
pressure 14 GPa with the one obtained from X-ray diffraction experiments \cite{Fun02}. The amplitude 
of the first peak of calculated PDF is a bit larger, that is perhaps a consequence of the GGA-PBE,
which usually overestimates the binding tendencies in liquid systems. However, the locations of 
the first and second maxima of PDF as well as the flat region between them are reasonably reproduced 
by the simulations.
\begin{figure}
\includegraphics[width=0.5\textwidth]{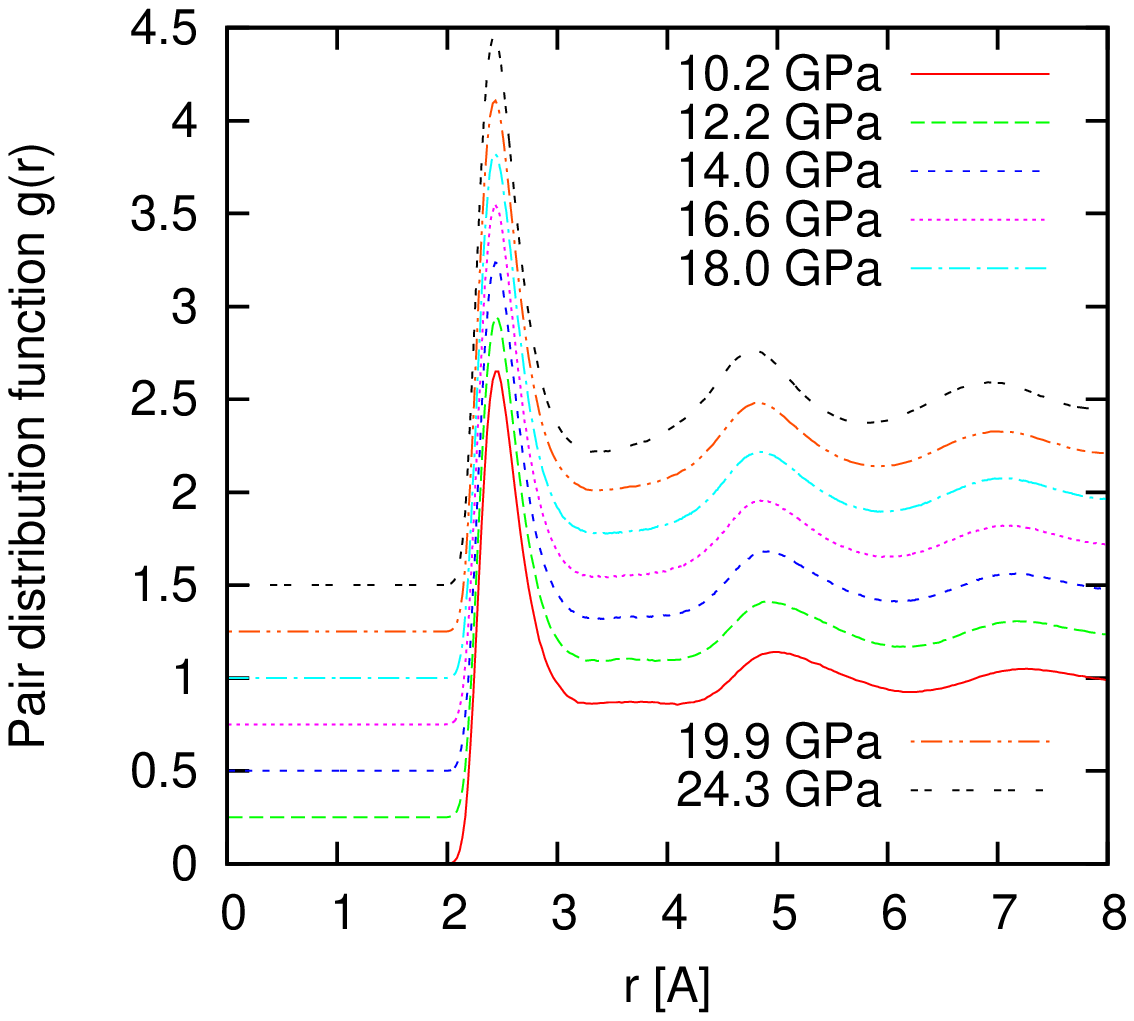}%
\includegraphics[width=0.5\textwidth]{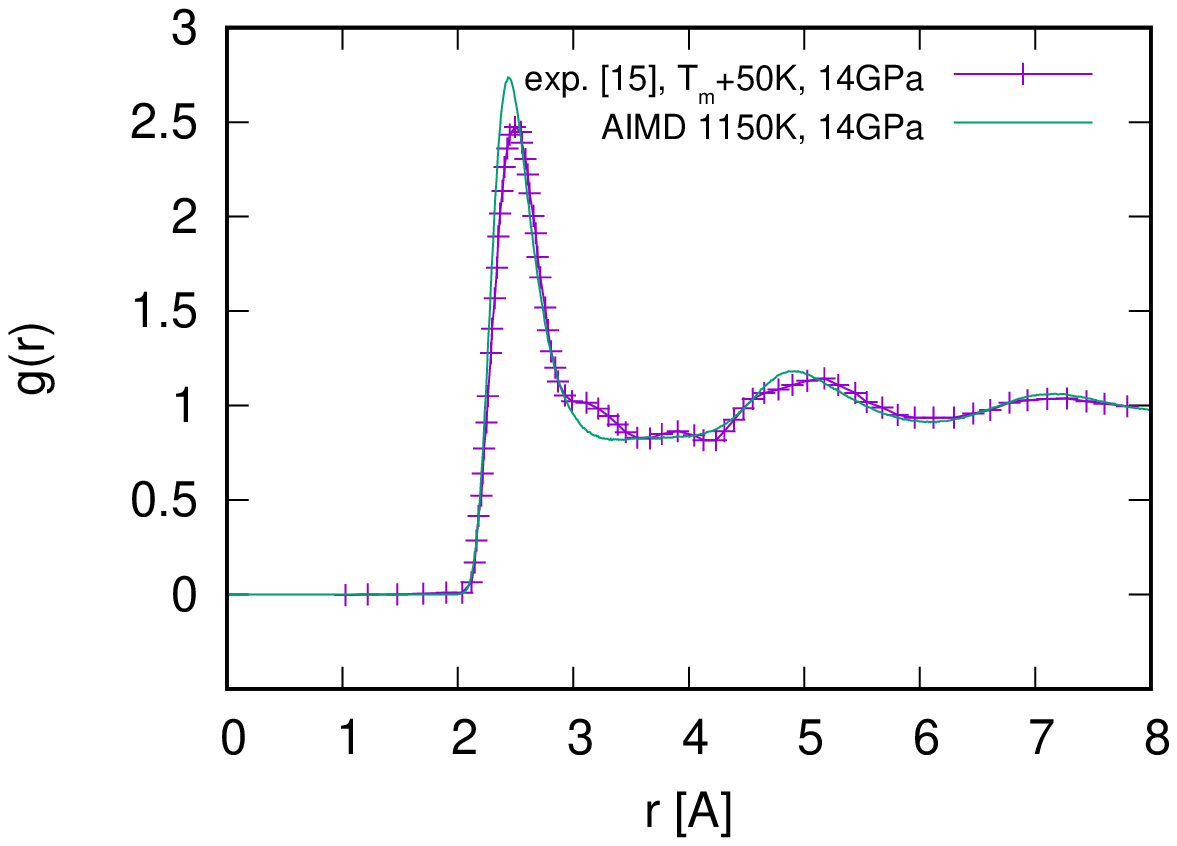}%
\caption{Pair distribution functions of liquid Si at T=1150K and pressure
range 10.2-24.3 GPa (a) and comparison with experimental data at 14~GPa taken 
from \cite{Fun02} (b).
}
\label{pdf}
\end{figure}

We tried to decompose the region around the first maximum of PDF into different shells in order 
to understand why the location of the first maximum of PDF does not change with the pressure increase
that is usually observed for liquid metals under pressure. The shell decompositions of the region 
under the first maximum of PDF is shown in Fig.3 for the lowest and highest studied pressures. We 
identified seven shells of neighbors under the first maximum, and for as one can see from Fig.3 
the main contribution comes from the three shells of the most bounded nearest neighbors within the 
distances up to 2.45\AA. These three
shells have larger amplitudes than the further shells and do not change their amplitudes with 
pressure increase. Perhaps the stability of the location of first maximum of PDF under pressure is 
the consequence of remaining in the liquid system covalent bonds, which cause in solid state 
tetrahedral structure of the nearest neighbors.
\begin{figure}
\includegraphics[width=0.5\textwidth,height=0.25\textheight]{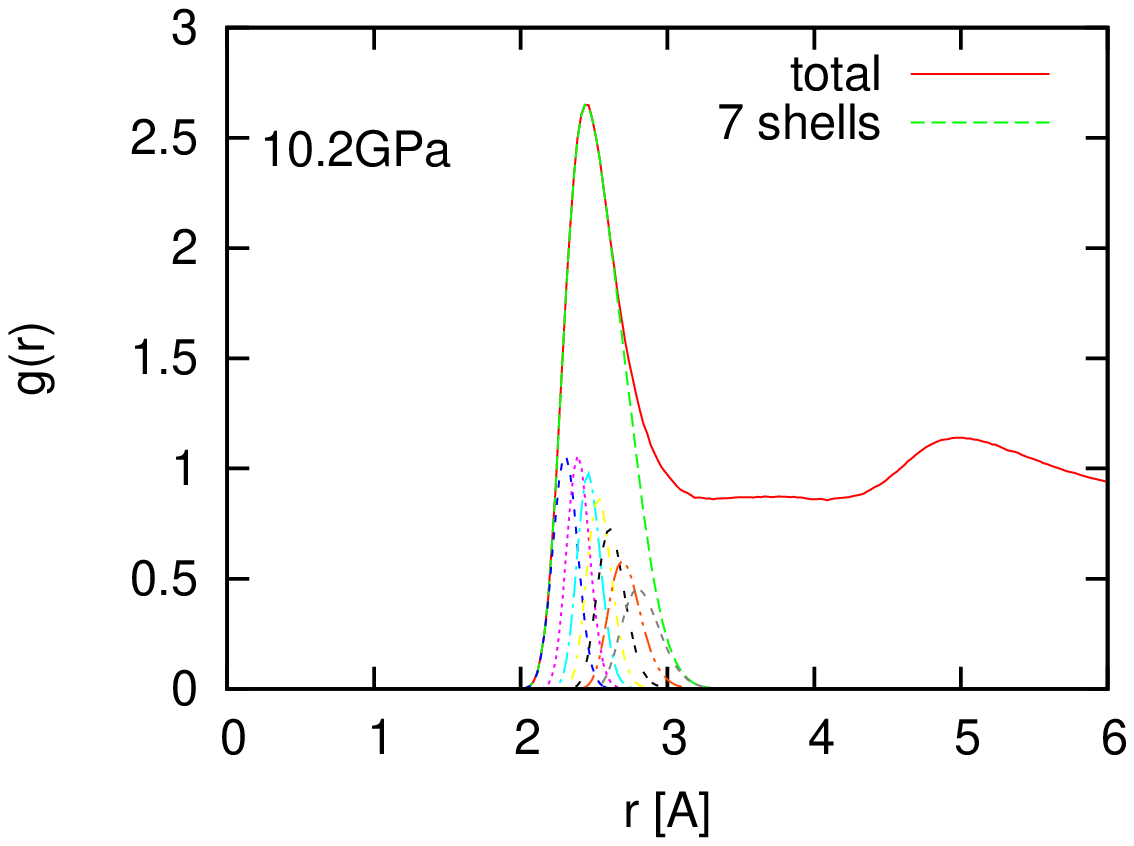}%
\includegraphics[width=0.5\textwidth,height=0.25\textheight]{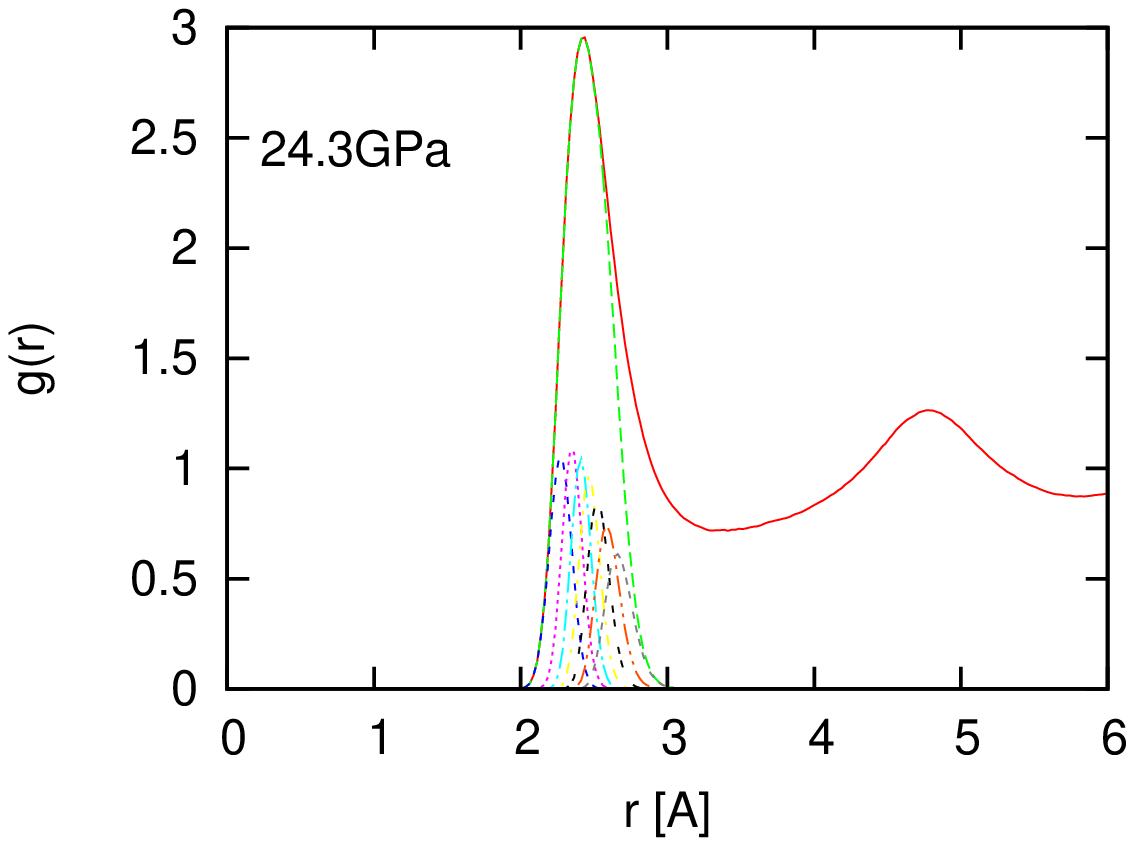}%
\caption{Decomposition of the first maximum of the pair distribution function
into shells of nearest neighbors st the smallest and highest studied pressures.
}
\label{shells}
\end{figure}

Further analysis of the nearest neighbors in the liquid Si we performed with the bond-angle 
distribution functions $g_3(cos{\theta})$, calculated within a cut-off radius of 
2.45\AA, that corresponds to the three shells of the most nearest neighbors. One can see 
in Fig.4 that the increase of the pressure results in reduction of triads of atoms with angles 
close to the tetrahedral one $\sim$109 degree with simultaneous increase of the triads with 
angles of the close-packed structure. 
\begin{figure}
\includegraphics[width=0.75\textwidth]{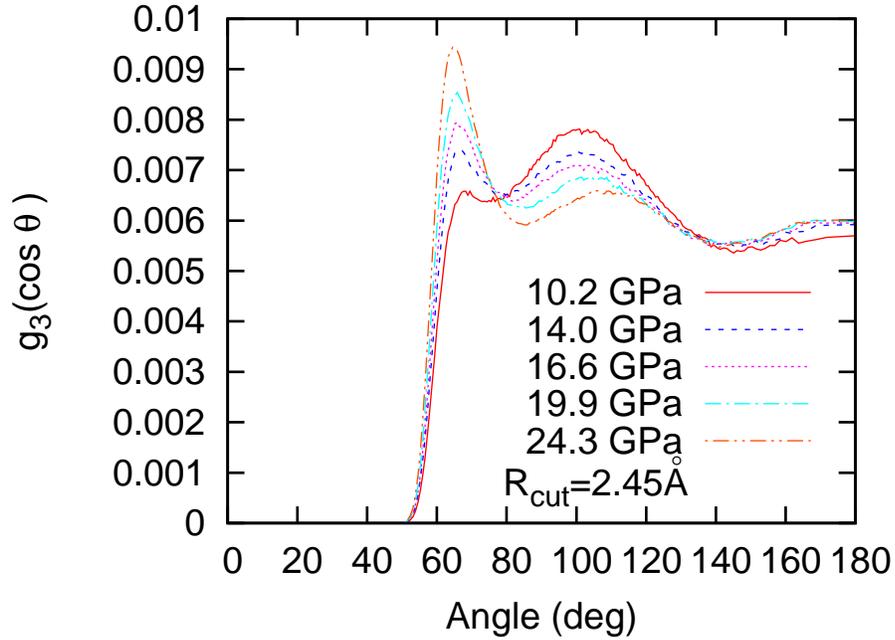}%
\caption{Pressure-induced change in three-body angular distribution functions 
$g_3(cos{\theta})$ for neighbors within the radius of 
2.45\AA, that corresponds to the three shells of the most nearest neighbors.
}
\label{bdf}
\end{figure}

The single-particle dynamics in liquid Si under pressure was studied via normalized 
velocity autocorrelation functions
$$
\psi(t)=\frac{\langle {\bf v}_i(t){\bf v}_i(0)\rangle_{i,t_0}}
{\langle{\bf v}_i(0){\bf v}_i(0)\rangle_{i,t_0}}~,
$$
where the averages were taken over the ensemble of particles as well as over different origins
$t_0$ of the time correlation functions. In Fig.5a one can see that for all the studied states 
the velocity autocorrelation functions show the cage effect, i.e. the negative values of $\psi(t)$ 
at small times which is caused by backscattering of particles due to 
collisions with nearest neighbors. The increase of pressure leads to a shift of this negative region 
towards smaller times that is naturally for liquids getting denser. The Fourier spectrum 
${\tilde Z}(\omega)$ of the 
velocity autocorrelation function in the high-frequency region reflects the vibrational density of 
states in the system, while its value at zero frequency is defined by the diffusion coefficient
$D$. In Fig.5b one can see that up to the pressures 16.6 GPa due to the high diffusivity the 
Fourier spectrum of the velocity autocorrelation function contains only one high-frequency peak, 
while at the highest pressure due to the drop of the diffusivity we clearly observed the two-peak 
structure of ${\tilde Z}(\omega)$, i.e. similar behavior as recently was observed for liquid 
Pb\cite{Bry19} and liquid Al\cite{Jak19} under pressure.
\begin{figure}
\includegraphics[width=0.5\textwidth,height=0.25\textheight]{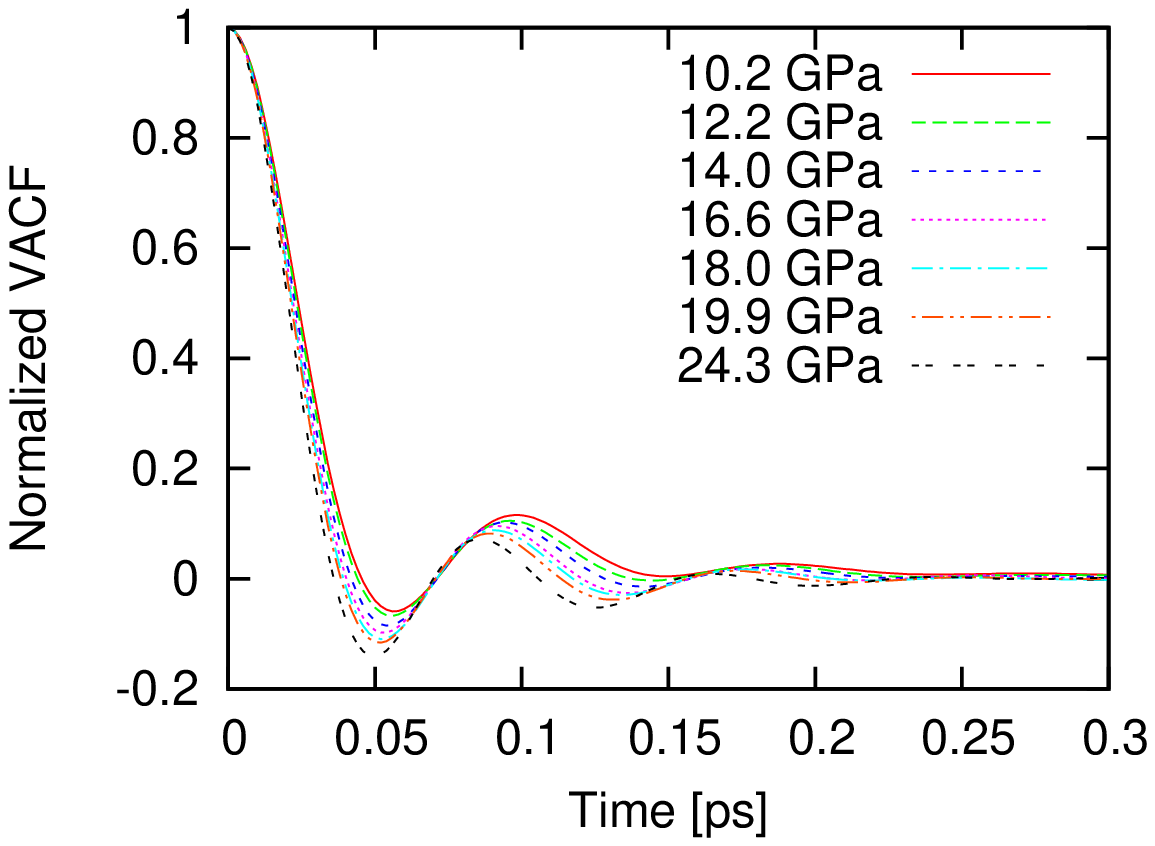}%
\includegraphics[width=0.5\textwidth,height=0.25\textheight]{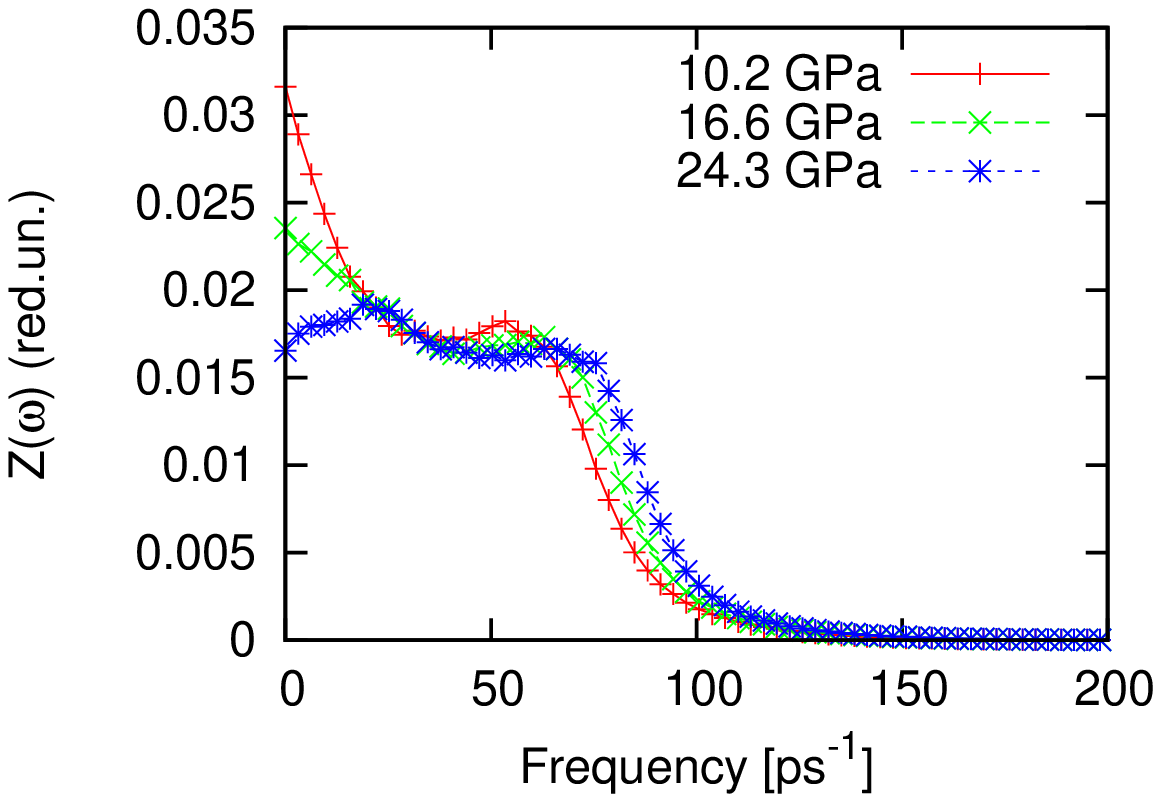}%
\caption{Velocity autocorrelation functions and their Fourier spectrum 
for liquid Si at T=1150K and three pressures.
}
\label{vacf}
\end{figure}

The diffusion coefficients, obtained from ${\tilde Z}(\omega=0)$, were verified by calculations 
of the long-time asymptotes of the mean square displacements $\langle R^2(t)\rangle$ \cite{Han}.
In Fig.6 we show the dependence of the two sets of data for diffusion coefficients of liquid Si 
along the isothermal line 1150~K on atomic volume. In \cite{Byl92} Bylander and Kleinman predicted 
a linear dependence of diffusivity vs atomic volume if there is no change in the effective 
radius of particles. Since in our Fig.6 the diffusivity fits very well to a linear dependence 
one can conclude that there is no change in the effective radius of Si atoms in the studied 
pressure range, and this is an argument for the absence of liquid-liquid transformations which 
can be expected in the studied region above the minimum of the melting line. Note, that we observed
the deviations from linear dependence of the diffusivity in liquid Rb at high pressures, where 
several other observations allowed to estimate the presence of structural transformation due to 
the change in effective radius of Rb atoms\cite{Bry13}. Also, a strong non-monotonic dependence 
of the diffusion coefficient was observed recently in fluid Hydrogen in the region of molecular-atomic
transition \cite{Ruo20}.
\begin{figure}
\includegraphics[width=0.75\textwidth]{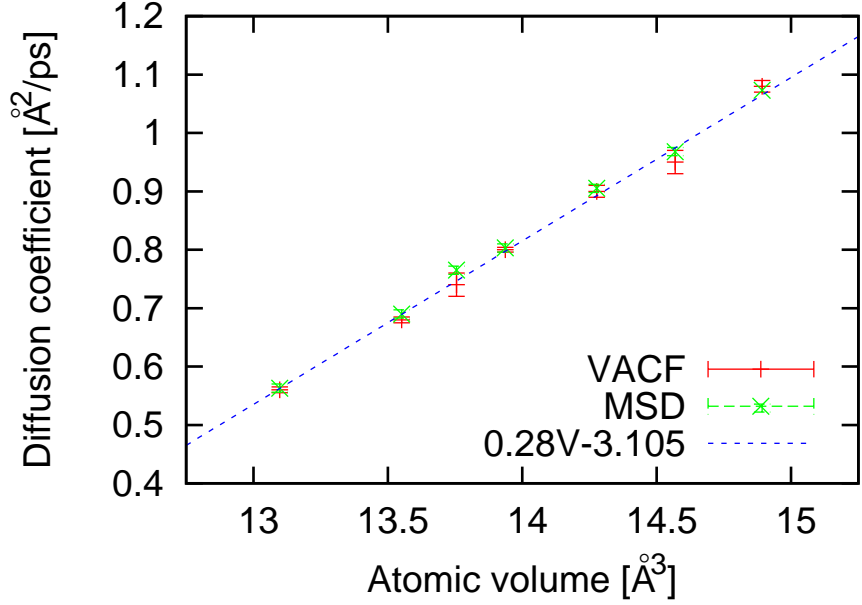}%
\caption{Diffusivity as a function of atomic volume in the pressure range 
10.2-24.3 GPa. The data obtained from the velocity autocorrelation functions (VACF) are 
shown by plus 
symbols with error bars, and the values from mean squre displacements (MSD) - by cross symbols with 
error bars.
 The straight line is a prediction by Kleinman and Bylander \cite{Byl92}.
}
\label{dif}
\end{figure}

Collective dynamics in liquid Si was studied via time-Fourier spectra of 
longitudinal (L) and transverse (T) time correlation functions
$$
F^{L/T}(k,t)=\langle J_{L/T}(k,t) J_{L/T}(-k,0)\rangle~,
$$
where we took the standard decomposition of the spatial Fourier components of 
total current into longitudinal and transverse components \cite{Han}. Note, that 
the local L-T cross-correlations (with the same wave number $k$) are zero as it should be
due to symmetry rules. The AIMD-derived L and T current-current time correlation functions were
used for calculations of their frequency spectra $C^{L/T}(k,\omega)$, peaks of which 
were treated as manifestations of collective excitations. In Fig.7 we show the dispersions
of L and T modes for four pressures, as they were obtained from peak positions 
of $C^{L/T}(k,\omega)$. We obtained very similar evolution of the L and T collective modes 
with pressure as were recently observed for liquid Pb\cite{Bry19}, liquid Al\cite{Jak19} 
liquid Na and In\cite{Bry20}:
for wave numbers in the second pseudo-Brillouin zone there exist two peaks on the transverse
spectral function $C^{T}(k,\omega)$. The transverse peak positions are shown by cross symbols 
with error bars in Fig.7. In general, the transverse branches in the second pseudo-Brillouin zone 
look almost flat and their mean frequency increases with pressure, and that is why we took the 
as some characteristic frequencies $\omega^T_{low}$
for the low-frequency T-band and $\omega^T_{high}$ for the high-frequency T-band in the second 
pseudo-Brillouin zone, similarly as it was done in \cite{Bry20}.
\begin{figure}
\includegraphics[width=0.5\textwidth,height=0.25\textheight]{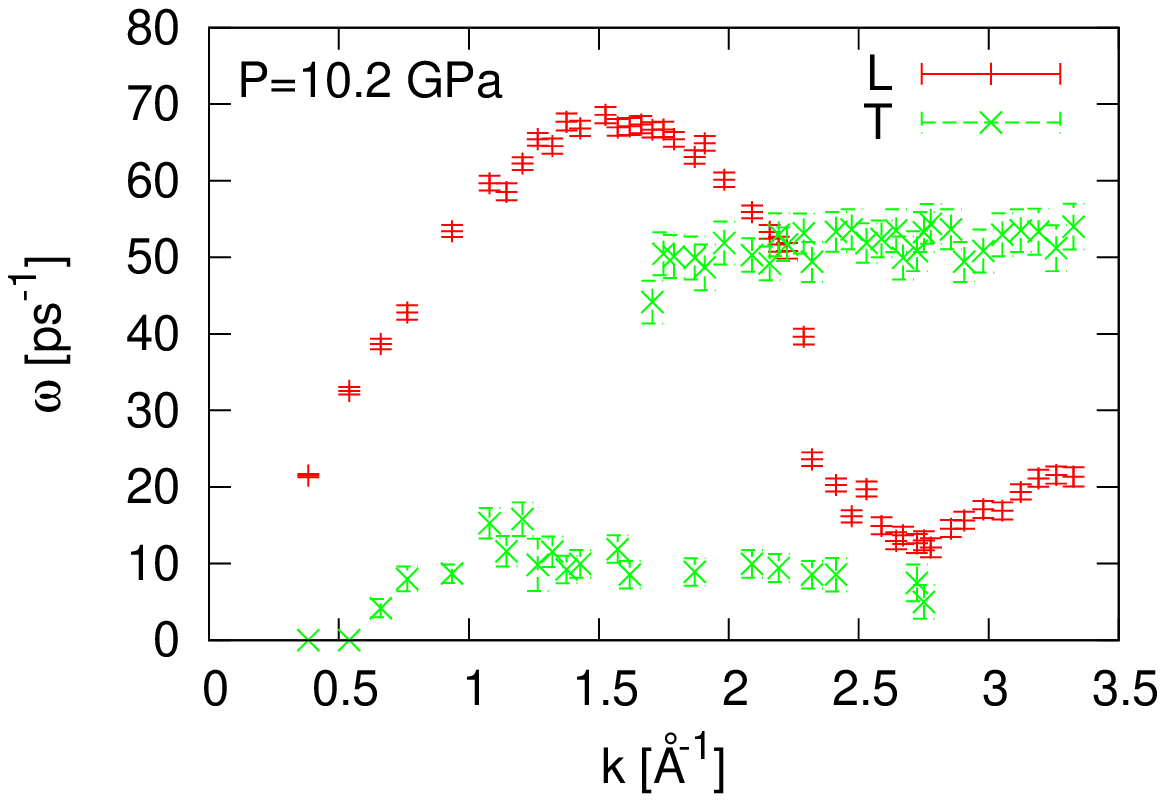}%
\includegraphics[width=0.5\textwidth,height=0.25\textheight]{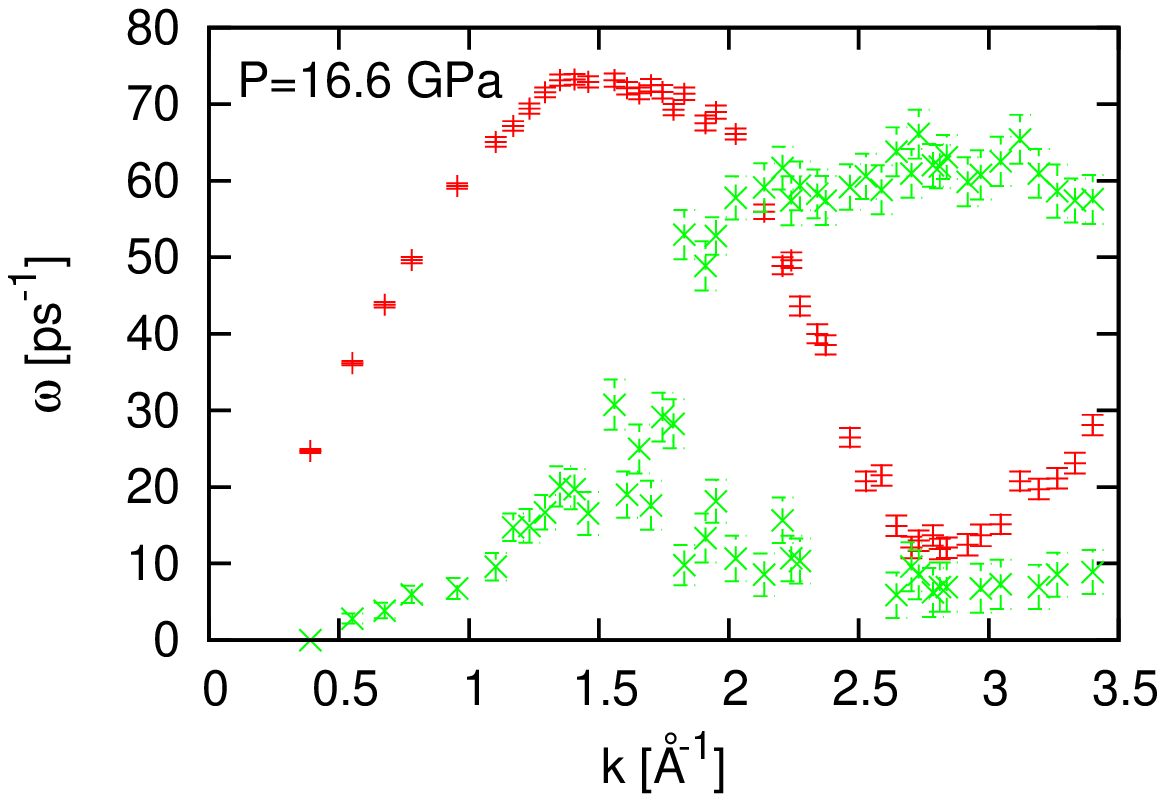}%

\includegraphics[width=0.5\textwidth,height=0.25\textheight]{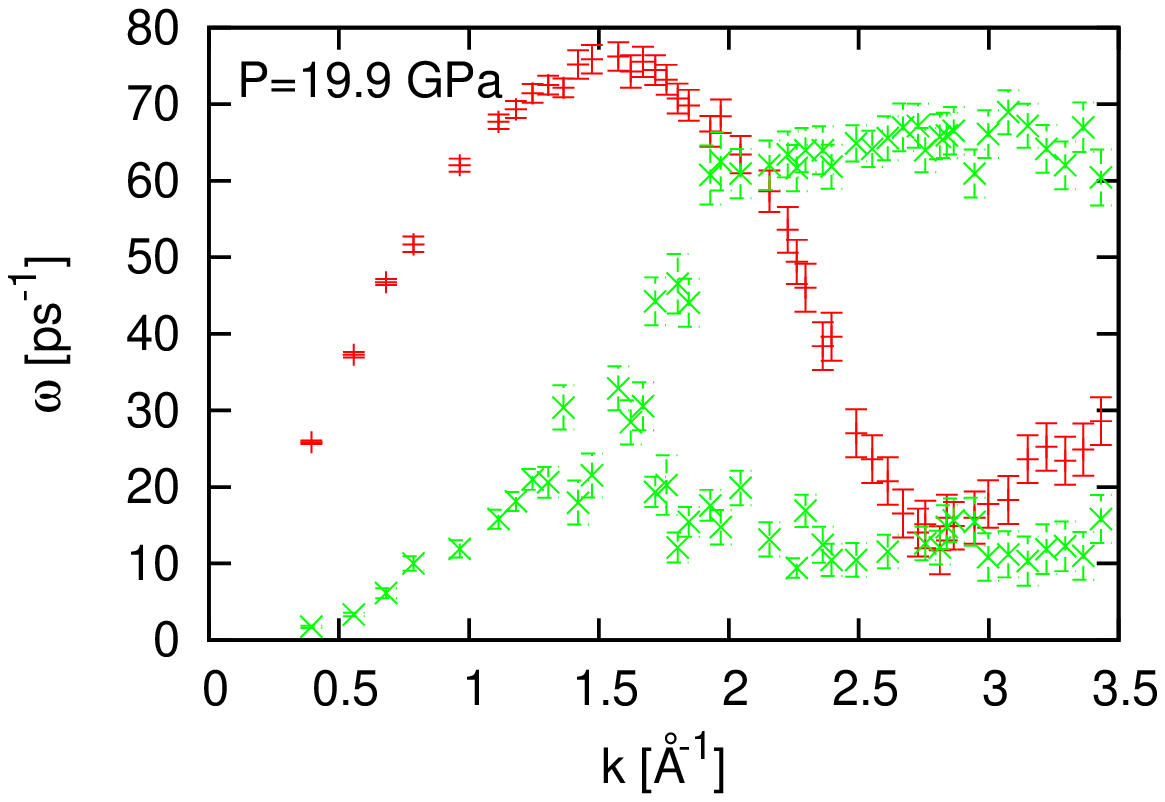}%
\includegraphics[width=0.5\textwidth,height=0.25\textheight]{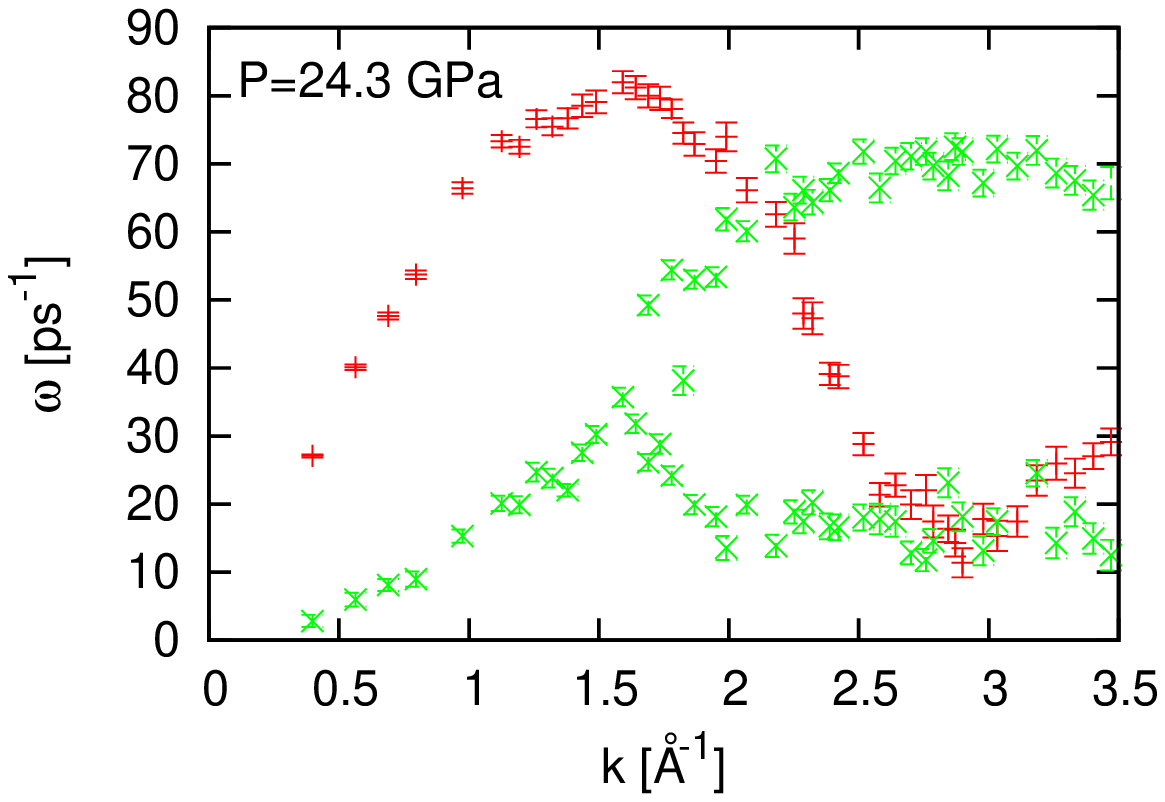}%
\caption{Evolution of dispersion of longitudinal (L) and transverse (T) 
collective excitations for liquid Si with pressure at T=1150K.
}
\label{disp}
\end{figure}

Similarly to the transverse case we took as some reference of the longitudinal dispersion 
the highest frequency of L-excitations $\omega^L_{max}$, which they usually took at the first 
pseudo-Brillouin zone boundary, i.e. Debye-like frequency. In Fig.8 we show the pressure dependence 
 of the three characteristic frequencies $\omega^L_{max}$, $\omega^T_{low}$ and $\omega^T_{high}$
and their correlation with the observed peak positions of the Fourier-spectra of velocity 
autocorrelation functions. One can see that for the two highest pressures, when the 
${\tilde Z}(\omega)$ contained two peaks, they both practically coincide with the frequencies  
$\omega^T_{low}$ and $\omega^T_{high}$ of the collective transverse current spectral functions.
For the smaller pressures the $\omega^T_{high}$ follows the dependence of the high-frequency 
peak of ${\tilde Z}(\omega)$.
By date it is not clear why the transverse current spectral functions show two-peak structure in 
the second pseudo-Brillouin zone. If this is due to the non-local L-T coupling (the local coupling 
is forbidden by the symmetry rules) then there should have been some similar features in the shape 
of longitudinal current spectral functions.
\begin{figure}
\includegraphics[width=0.75\textwidth]{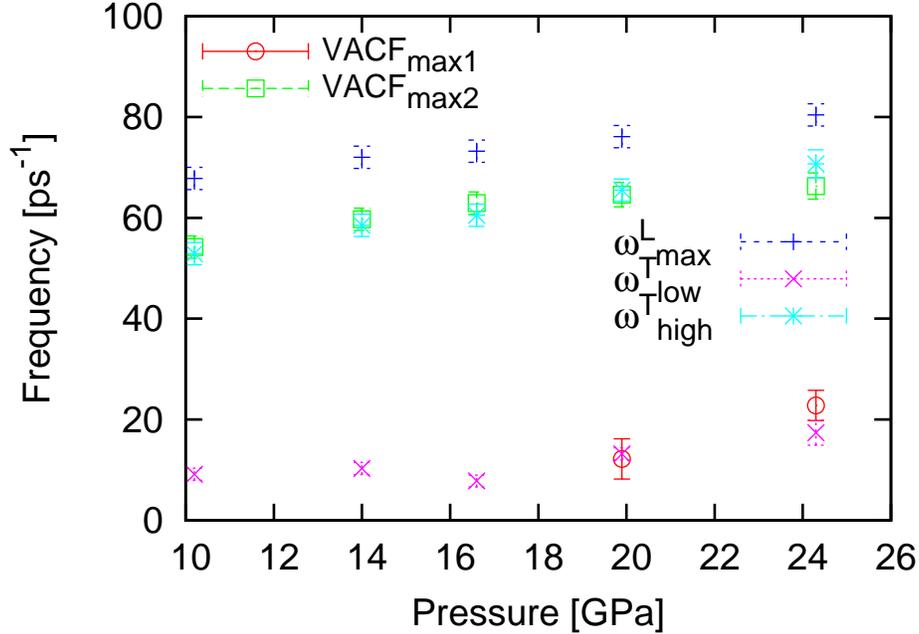}%
\caption{Correlation in peak locations of Fourier-spectra of velocity autocorrelation
functions and highest frequency of longitudinal dispersion (Debye-like frequency) $\omega^L_{max}$
and flat regions of two transverse branches in the second pseudo-Brillouin zone $\omega^T_{low}$ and
$\omega^T_{high}$ in the pressure range 10.2-24.3 GPa. 
}
\label{freq}
\end{figure}

\section{Conclusion}

We presented our results of AIMD simulations for liquid Si along the isothermal line 1150~K 
which is above the minimum of the melting line of Si. Neither the studied structural quantities 
nor the pressure dependence of diffusion coefficient showed any feature for pressures above the
minimum of the melting line, that makes evidence of the absence of liquid-liquid transformations 
in liquid Si in this region.

The Fourier spectra of single-particle velocity autocorrelation functions ${\tilde Z}(\omega)$ 
and of collective transverse current-current correlation functions $C^{T}(k,\omega)$ (for wave 
numbers $k$ in the second pseudo-Brillouin zone) show very similar locations of their peaks.
This is in line with the recent results for liquid Pb\cite{Bry19}, liquid Al\cite{Jak19}, 
liquid Na and In \cite{Bry20}. We cannot so far identify the origin of the high-frequency 
peak of $C^{T}(k,\omega)$, which appears in the second pseudo-Brillouin zone. From the point 
of view of the manifestation of the L-modes in the spectra of transverse excitations due to 
broken spherical symmetry on the microscopic scale of nearest neighbors the local coupling 
(with the same wave number $k$) is prohibited for averages over all possible configurations, 
while the non-local mode coupling approach can be more useful in this case, however so far 
it was applied very rarely \cite{Rio17}.

{\it Acknowledgments}
The calculations have been performed using the ab-initio total-energy
and molecular dynamics program VASP (Vienna ab-initio simulation program)
developed at the Institute f\"ur Materialphysik of the Universit\"at Wien
\cite{Kre93,Kre96,Kre96b}. TD and TB were supported by the 
joint French-Ukrainian "Dnipro" project M-70/2019. APS was supported by the 
CNRS PISC project "Ab initio simulations of
structural and dynamic features of complex and molecular fluids of geophysical 
interest".
TD benefited from new courses on AIMD of the ICMP PhD program developed in frames of
DocHub/Erasmus+ project 574064-EPP-1-2016-1-LT-EPPKA2-CBHE-SP.


%
%
\end{document}